\documentclass[a4paper]{jpconf}
\usepackage{graphicx}
\begin{document}
\title{Open Source High Fidelity Modeling of a Type 5 Wind Turbine Drivetrain for Grid Integration \footnote{This manuscript is originally submitted to Journal of Physics. }}

\author{Tanveer Hussain$^1$, Juan Gallego-Calderon$^1$, S M Shafiul Alam$^1$}

\address{$^1$ Power and Energy Systems Department, Idaho National Laboratory, 1955 N Fremont Ave, Idaho Falls, ID 83415, United States.}

\ead{tanveer.hussain@inl.gov}

\begin{abstract}
The increasing integration of renewable energy resources in evolving bulk power system (BPS) is impacting the system inertia. Type-5 wind turbine generation has the potential to behave like a traditional synchronous generator and can help improve system inertia. Hydraulic torque converter (TC) and gearbox with torque limiting feature are integral parts of a Type-5 wind turbine unit. High fidelity model of Type-5 wind turbine including these core components is not openly and widely available for grid integration and transient stability studies. This hinders appropriate assessment of Type-5 wind power plant's contribution to bulk grid resilience. This work develops a TC model based on those generally used in automobile's transmission system. Moreover, the concept of torsional coupling is leveraged to integrate the TC and gearbox system dynamics. The entire integrated model will be open sourced and publicly available for grid integration studies.
\end{abstract}

\section{Introduction}
The share of variable renewable energy (VRE) sources like wind and solar in the bulk power system are expected to grow significantly in the next two to three decades. For example, the United States (U.S.) has ambitious targets of powering the country with 100\% renewables by 2035. In terms of offshore wind, the most immediate objective is to deploy 30 GW of offshore wind by 2030 with a pathway to 110 GW by 2050 \cite{walter_musial_offshore_2022}.  This makes wind power one of the fastest growing generation technologies, becoming a major contributor to the energy supply of many countries. Denmark is leading the way with over 40\% of the total generation as of 2021, while the European Union (EU) as a whole depends on over 10\% of wind penetration. In contrast, land-based wind supplied the U.S. with 9.1\% of total electricity generation for the same year \cite{ryan_wiser_land-based_2022}.

The trend towards 100\% renewables will impose several challenges in a synchronous generator based power system. Inverter based resources, like VREs, have traditionally lacked capabilities to provide ancillary services, such as inertia support in the bulk power system and grid stability, in the same way as traditional synchronous generators (e.g., hydropower plants) have contributed. Yet, the vast majority of currently installed wind energy capacity is distributed between Type-3 and Type-4 turbines. In both configurations, the use of power converters is needed to operate the wind turbine in variable speed configuration, to isolate the generator from the grid, and for complying to grid codes \cite{gevorgian_grid-forming_2022}. Conversely, a rarely explored configuration, so-called Type-5 wind turbine, is believed to gain traction so that the high penetration of wind energy can support the aforementioned grid services.

The impact of variable-speed wind turbines using synchronous generators has been lightly investigated. A major contribution was produced by \cite{muller_h_grid_2006}, where a detailed model of a 2 MW drivetrain composed of a gearbox, hydrodynamic torque converter, and a synchronous generator, was developed in DIgSILENT with the purpose to investigate the impact on grid stability. The study amplified the individual turbine model to a 50 MW wind farm under turbulent wind, and found several advantages on using this configuration: the wind farm was able to provide voltage support and provide stability by increasing the short-circuit level; the concept is able to operate in islanded mode; and, an outlook towards offshore wind farms was provided where they can be connected directly to high-voltage-direct-current (HVDC) systems using conventional thyristor technology. 

As mentioned above, a torque converter is needed to operate a turbine in synchronous mode. Yet, the publicly available tools and research in the detailed modeling of the drivetrain is sparse. The model developed in \cite{muller_h_grid_2006} is not publicly available and it is the only paper in the literature looking specifically into the dynamics of a Type-5 drivetrain, for example. For decades, the automotive industry has worked in torque converters and the literature in this field is abundant \cite{hrovat_bond_1985, yang_li_modellingmeasurementstransienttorqueconverterchar_2016pdf_2016, deur_analysis_2002}. In this paper, we describe the modeling of the torque converter in the context of wind energy, along with its coupling with a gearbox and synchronous generator. The main objective of this work is to develop an open-source dynamic model of a Type-5 wind turbine for research and development purpose. The following section, describes the torque converter modeling in detail, along with couplers used to integrate other parts of the drivetrain (gearbox and generator). First, the established models of coupler and gearbox are briefly discussed. The introduction to torque converter (TC) dynamic model then follows according to the automobile literature. The automobile-grade TC model is then validated. The scaling of a validated automobile-grade TC to Type-5 wind turbine is then described. Finally, Type-5 TC's initialization and sensitivity analysis is carried out to develop a speed governing system.  


\section{Modeling}

\begin{figure}[h] 
\includegraphics[width=\textwidth]{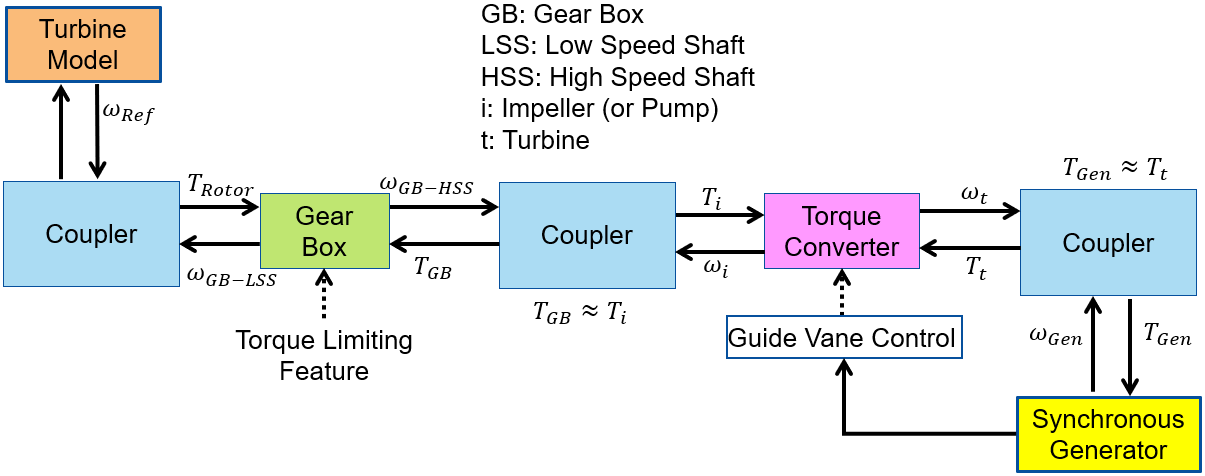}\hspace{2pc}%
\caption{\label{fig:integratedModel}Integrated design block diagram. Here, the "turbine model" can be represented using open-source aerolelastic codes such as HAWC2 or FAST.}
\end{figure}

Unlike Type-3 and Type-4 wind turbines, the Type-5 configuration uses a fixed-speed synchronous generators that is directly connected to the grid (i.e., no power converter is required). In synchronous operation, the generator speed is dictated by the grid frequency (i.e. 50 Hz in EU, and 60 Hz in the U.S.), instead of the wind turbine variable speed. This is achieved thanks to a torque converter (TC) placed between the gearbox and generator (Figure \ref{fig:integratedModel}). The purpose of the TC is to convert the variable speed "seen" in the wind turbine rotor to a fixed speed, thanks to either a hydro-static or hydraulic system that dissipates the torque. The TC has two rotating parts and one static part, i.e., the impeller/pump, the turbine, and the guide vane, respectively. The impeller is connected to the high-speed stage of the gearbox whereas the turbine is connected to the generator. The guide vane, which is placed between the impeller and the turbine, controls the returning fluid from the turbine to the impeller. To maintain a torque balance in the system, each major component (i.e., gearbox, TC, and generator) is connected using a torsional coupler (\ref{eq:coupler}). The gearbox is modeled using a lumped parameter approach with torsional and transnational degrees of freedom. The interaction in the gear mesh and the bearings is modeled using linear springs \cite{gallego-calderon_juan_electromechanical_2015}. The impeller and turbine torques are the inputs to the TC dynamics model whereas impeller and turbine speeds are the outputs of the model \cite{hrovat_bond_1985, deur_analysis_2002}. The guide variable $\alpha_s$ controls the position of the guide vane. The guide vane redirects the fluid from the turbine to the impeller in such a way that the turbine speed of the TC remains constant.  

The integrated design (Figure \ref{fig:integratedModel}) approach in this project, uses coupler blocks (\ref{eq:coupler} to maintain the torque balance in the free-free system. This ensures that the conservation of energy principle is respected in the system of equations, and integrates the intermediate shafts torsional flexibility in the simulation. This is particularly important when coupling the Simulink model with an external aeroelastic tool in a co-simulation environment.

The gearbox is modeled as a lumped-parameter system that includes translational and torsional degrees of freedom. This means that it is possible to compute the mechanical loading due to radial forces in the bearings. The torsional component transfer the torque from the rotor side to the impeller side of the TC, and vice-versa. This capability allows for future studies on characterizing the torsional loading in the gearbox due to transient events. Moreover, the translational component provides the displacements in the vertical and horizontal directions of the bearings (e.g., planetary) due to the wind turbine dynamics. The model was developed using an object-oriented approach and validated previously by \cite{gallego-calderon_juan_assessment_2015, gallego-calderon_juan_electromechanical_2015}.

\begin{equation} \label{eq1}
    T_r = K_s \int (\omega_2 - \omega_1)d\omega + C_s(\omega_2 - \omega_1)
    \label{eq:coupler}
\end{equation}


\subsection{Torque Converter}
\begin{figure}[h] 
\includegraphics[width=\textwidth]{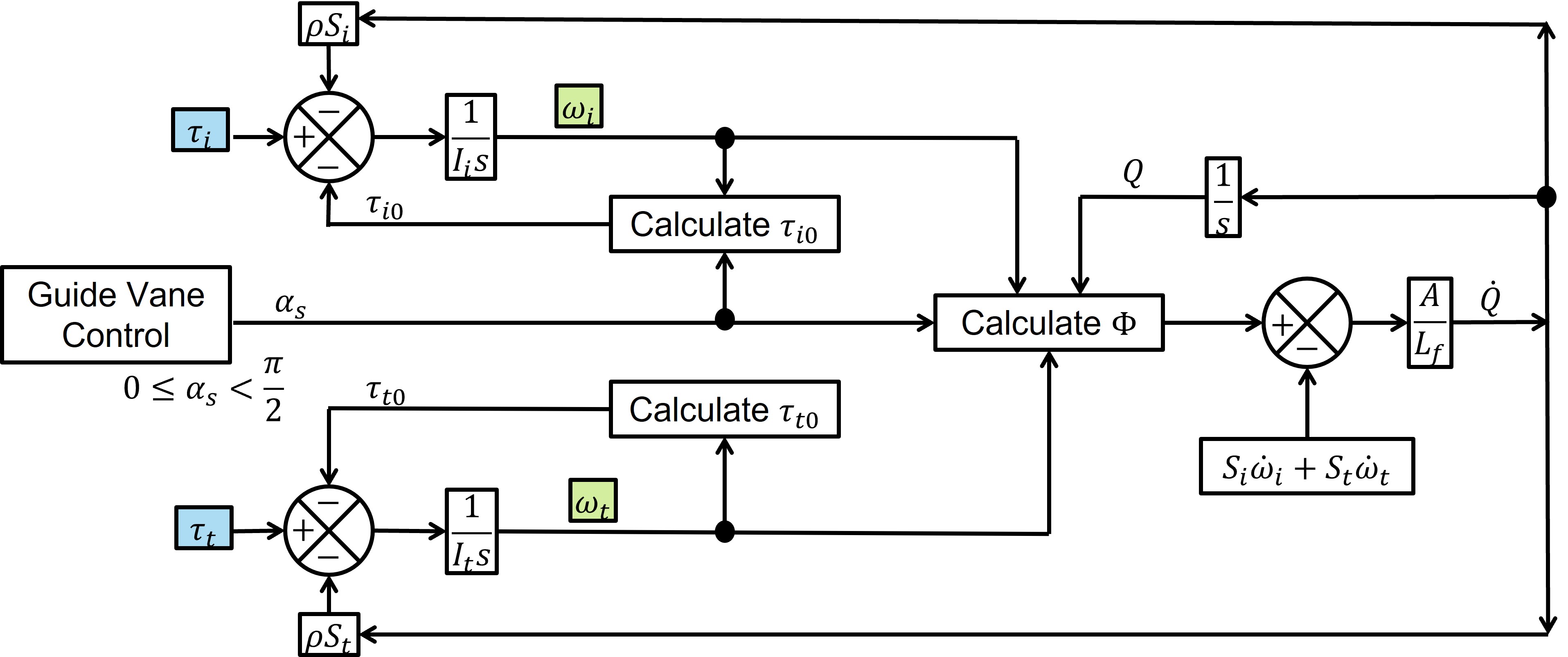}\hspace{2pc}%
\caption{\label{TCfig}Block diagram of torque converter model.}
\end{figure}

The dynamics of a torque converter (TC) is defined by the following set of three equations \cite{hrovat_bond_1985,deur_analysis_2002}, and is shown in Fig. \ref{TCfig}

\begin{equation} \label{eq1}
    I_t\dot{\omega_t} + \rho AS_t\dot{V} = \tau_t - \tau_{t0}(V,\omega_i,\omega_t)
    \label{eq:tto}
\end{equation}

\begin{equation}\label{eq2}
    I_i\dot{\omega_i} + \rho AS_i\dot{V} = \tau_i - \tau_{i0}(V,\omega_i,\alpha_s)
    \label{eq:tio}
\end{equation}

\begin{equation} \label{eq3}
    S_i\dot{\omega_i} + S_t\dot{\omega_t} + L_f\dot{V} = \Phi(V,\omega_i,\omega_t,\alpha_s)
    \label{eq:Phi}
\end{equation}

Here, $\omega_i$ and $\omega_t$ represent impeller angular speed and turbine angular speed, respectively. Speed ratio is defined as $\nu = \omega_t/\omega_i$. $\tau_i$  and $\tau_t$ represent impeller torque and turbine torque, respectively. $A$ represents the cross-sectional area perpendicular to the volume flow. $\rho$ is the density of the working fluid. $I_i$ and $I_t$ are impeller and turbine's moment of inertia, respectively. $S_i$ and $S_t$ are characteristic area constants, respectively. $V$ is the fluid velocity and $ L_f$ is equivalent fluid inertia length. At the steady-state condition, $\tau_t = \tau_{t0}(V,\omega_i,\omega_t)$, $\tau_i = \tau_{i0}(V,\omega_i,\alpha_s)$, and $\Phi(V,\omega_i,\omega_t,\alpha_s) = 0$.  The variables $\tau_{i0}$  and $\tau_{t0}$ in equations \ref{eq1} and \ref{eq2} represent the steady-state values of impeller and turbine torques, respectively, and are given by,
\begin{equation}
    \tau_{i0} = \rho Q[\omega_iR_i^2 + V(R_i\;tan\alpha_i -R_s\;tan\alpha_s )]
\end{equation}

\begin{equation}
    \tau_{t0} = \rho Q[\omega_tR_t^2-\omega_iR_i^2 + V(R_t\;tan\alpha_t -R_i\;tan\alpha_i )]
\end{equation}

where, Q is the axial torus volume flow and is given by, $Q = V\times A$. $R_i$, $R_t$, and $R_s$ are impeller, turbine, and stator exit radius, respectively. Similarly, $\alpha_i$, $\alpha_t$, and $\alpha_s$ are impeller, turbine, and stator exit angles, respectively.

It is assumed that the stator is fixed and will not rotate across different $\nu$ values. $\omega_t$ that drives the calculation of $\nu$ for Type 5 Wind turbine speed range is given by,

\begin{equation}
    \omega_t [rad/sec] = \frac{N\times120\times\pi}{3600}
\end{equation}
Here, $N$[rpm] represents the rated speed of the synchronous generator. Stator exit angle is presented by $\alpha_s$, and its usage for stator guide vane control is explained in later section. The variable $\Phi$ in equation \ref{eq3} is given by,
\begin{equation}
    \Phi = R_i^2 \omega_i^2+ R_t^2 \omega_t^2   - R_i^2 \omega_t\omega_i+\omega_iV(R_i\;tan\alpha_i - R_s\;tan\alpha_s ) + \omega_tV(R_t\;tan\alpha_t - R_i\;tan\alpha_i ) - \psi 
\end{equation}
where $\psi$ is given by,

\begin{equation} \label{eq8}
\psi  =  0.5[C_{sh,i}V_{sh,i}^2 +C_{sh,t}V_{sh,t}^2+C_{sh,s}V_{sh,s}^2 + f (V_i^{*2}+V_t^{*2}+V_s^{*2})] 
\end{equation}
where, $f$ is frictional loss coefficient.   $C_{sh,i}$, $C_{sh,t}$, and $C_{sh,s}$ are impeller, turbine, and stator's shock loss coefficients. Similarly, $V_{sh,i}$, $V_{sh,t}$,and $V_{sh,s}$ are impeller, turbine, and stator's shock velocities and are given by, 
\begin{equation}
 V_{sh,i}= -R_s \omega_i + V(tan\alpha_s-\;tan\alpha_{i}')
\end{equation}

\begin{equation} 
 V_{sh,t}= R_i( \omega_i- \omega_t )+ V(tan\alpha_i-\;tan\alpha_{t}')
\end{equation}

\begin{equation} 
 V_{sh,i}= R_t \omega_t + V(tan\alpha_t-\;tan\alpha_{s}')
\end{equation}
where, $\alpha_i'$, $\alpha_t'$, and $\alpha_s'$ are impeller, turbine, and stator inlet angles, respectively.

$V^*$ in equation \ref{eq8} is the fluid velocity relative to blades, and is given by,
\begin{equation} 
V_i^{*2} = V^2\;sec\;\alpha_i^2
\end{equation}

\begin{equation} 
V_t^{*2} = V^2\;sec\;\alpha_t^2
\end{equation}

\begin{equation}
V_s^{*2} = V^2\;sec\;\alpha_s^2
\end{equation}

\subsubsection{Model Validation Using Honda CRV Values from \cite{asl_math-based_2014}}
\begin{figure}[h!]
\centering
\includegraphics[width=0.5\textwidth]{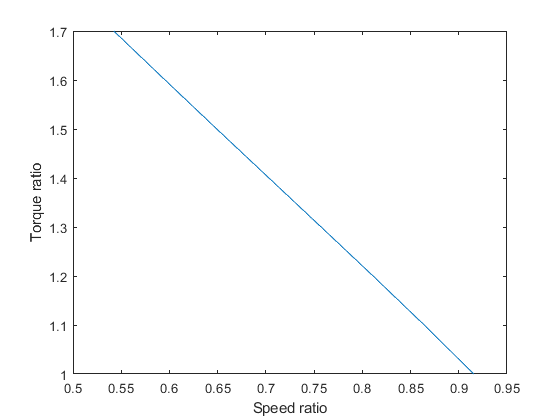}\hspace{2pc}%
\caption{\label{Ratio}Torque ratio vs. speed ratio \cite{asl_math-based_2014}.}
\end{figure}

:
The proposed torque converter model is validated on auto-transmission scale using Honda CRV values from \cite{asl_math-based_2014}. The simulation results are based on steady state calculations. Table \ref{tab:my-table} shows the values of the torque converter parameters.Using these parameters, Fig. \ref{Ratio} shows the characteristic plot for the torque ratio ($\tau_t/\tau_i$ ) vs. the speed ratio, $\nu$.

\begin{table}[h!]
\centering
\caption{Nominal values of the torque converter parameters of Honda CRV \cite{asl_math-based_2014}}
\label{tab:my-table}
\begin{tabular}{|l|l|l|l|}
\hline
Fluid density ($\rho$)         & 840 kg/$m^3$  & Fluid inertia length ($L_f$)       & 0.2594 m   \\ \hline
Flow area (A)             & 0.0107 $m^2$  & Impeller inertia ($I_i$)           & 0.092 kg $m^2$  \\ \hline
Impeller radius ($R_i$)      & 0.0991 m     & Turbine inertia ($I_t$)            & 0.026 kg$m^2$  \\ \hline
Turbine radius ($R_t$)       & 0.0735 m    & Stator inertia ($I_s$)             & 0.012 kg$m^2$  \\ \hline
Stator radius ($R_s$)        & 0.0665 m    &  Impeller Shock loss coefficient ($C_{sh}$)    & 1.011      \\ \hline
Impeller exit angle ($\alpha_i$)  & 16.21 $ ^{\circ} $ & Turbine Shock loss coefficient ($C_{sh}$)    & 1.8         \\ \hline
Turbine exit angle ($\alpha_t$)   & -53.14 $ ^{\circ} $ &  Stator Shock loss coefficient ($C_{sh}$)    & 0.773      \\ \hline
Stator exit angle ($\alpha_s$)    & 55.62 $ ^{\circ} $     & Impeller design constant ($S_i$)   & -0.001 $m^2$   \\ \hline
Impeller inlet angle ($\alpha_i'$) & -40.7 $ ^{\circ} $  & Turbine design constant ($S_t$)    & -0.00002 $m^2$ \\ \hline
Turbine inlet angle $\alpha_t'$)  & 59.19 $ ^{\circ} $  & Stator design constant ($S_s$)     & 0.002 $m^2$    \\ \hline
 Stator inlet angle ($\alpha_s'$)         & 60.36 $ ^{\circ} $ & Frictional loss coefficient (f) & 0.197  \\ \hline
\end{tabular}
\end{table}

\begin{figure}[h]
\centering
\includegraphics[width=0.8\textwidth]{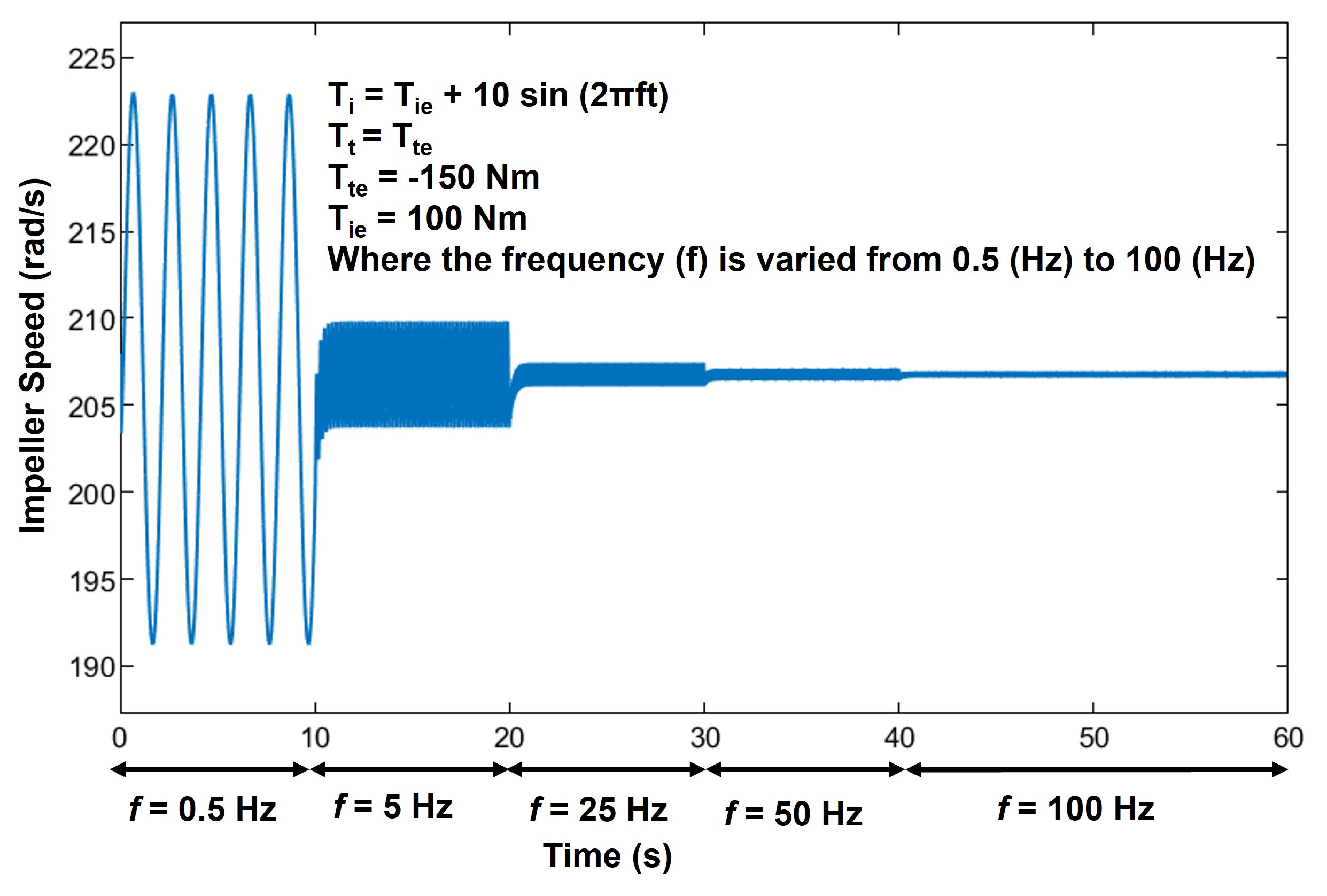}\hspace{2pc}%
\caption{\label{ImpellerFreqResponse}Nonlinear torque converter model response for impeller side when frequency is varied from 0.5-100 Hz \cite{asl_math-based_2014}.}
\end{figure}

\begin{figure}[h]
\centering
\includegraphics[width=0.8\textwidth]{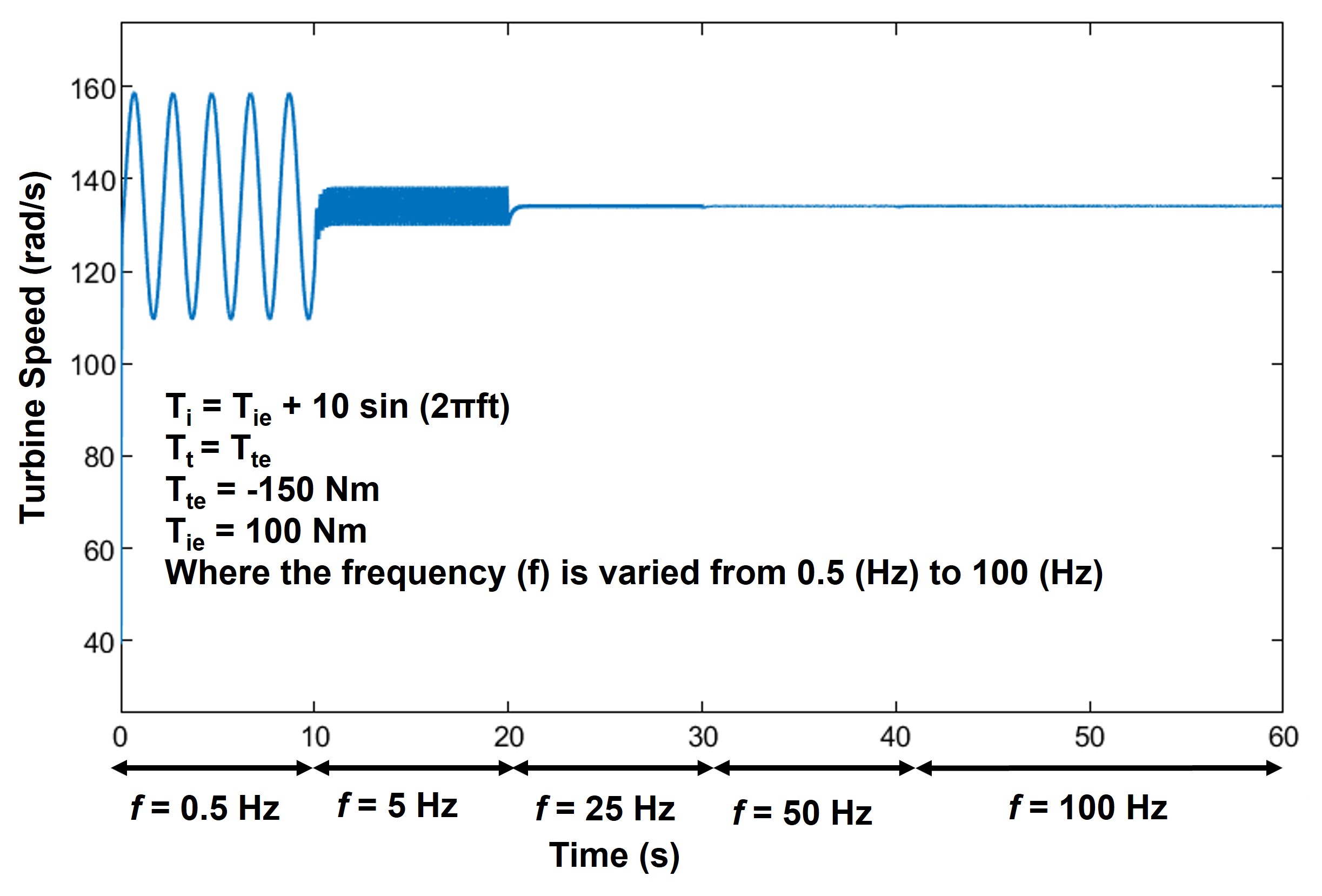}\hspace{2pc}%
\caption{\label{TurbineFreqResponse}Nonlinear torque converter model response for turbine side when frequency is varied from 0.5-100 Hz \cite{asl_math-based_2014}}
\end{figure}

One of the advantages of using a torque converter in a type 5 wind turbine is its' capability to filter disturbances introduced from impeller side. Figures \ref{ImpellerFreqResponse} and \ref{TurbineFreqResponse} shows the torque converter frequency response analysis. The torque converter is working as a low pass filter to damp disturbances which are transferred from the impeller side to the turbine side at high frequencies. For validation, we tried to mimic the Figure 20 from \cite{asl_math-based_2014}. The impeller torque is subjected to a disturbance in a wide range of frequencies, i.e., the impeller torque is given as a combination of a nominal value and disturbances $\tau_i = \tau_{ie} + 10\;sin(2\pi ft)$ where the frequency ($f$) is varied from 0.5 ($Hz$) to 100 ($Hz$), and the turbine torque $\tau_{t}$ is assumed constant, i.e., $\tau_{te}$. The values for $\tau_{ie}$ and $\tau_{te}$ used for this case study are 100 $Nm$ and -150 $Nm$, respectively. The plots of the impeller speed ($\omega_i$ ) and the turbine speed ($\omega_t$), Figures \ref{ImpellerFreqResponse} and \ref{TurbineFreqResponse} , show the damping characteristics of the torque converter to filter high frequency disturbances from impeller to turbine side.

\subsection{Parameter Scaling to Type 5 Wind Turbine}
The TC model developed and validated according to Honda CRV specifications is needed to scale up to match torque and speed requirements of a Type 5 wind turbine. First, the turbine torque $\tau_t$ is defined as follows,

\begin{equation}
    \tau_t= \Bigg\{\begin{array}{ccl} -\tau_{rated}& ;0.87 \leq \nu \leq 1  \\
-\frac{\tau_{rated}}{\nu^2}& ; 1 < \nu \leq 1.67
\end{array}
    \label{eq:TvsSR}
\end{equation}
Here, $\tau_{rated} = \frac{30\times P_{rated}}{\pi\times N}$ [N-m]. The operating range of $\nu$ and $P_{rated}$ follow DeWind D8.2 type 5 wind turbine specifications \cite{Dewind}. The goal is to a) scale up the geometric parameters such as length, radius and flow area, and b) adjust impeller and turbine exit angles to ensure minimum absolute value of $\Phi$ at the steady state condition of unity torque ratio (i.e., $\tau_i = |\tau_t|$), and unity speed ratio. This will enable Type 5 TC  initialization at synchronous speed and rated power. Let, $K$ denote the  geometric parameter amplification factor such that,
\begin{equation}
    \mathrm{Length}(\mathrm{Type5}) = K\times\mathrm{Length}(\mathrm{HondaCRV})
\end{equation}

\begin{equation}
    \mathrm{Radius}(\mathrm{Type5}) = K\times\mathrm{Radius}(\mathrm{HondaCRV})
\end{equation}

\begin{equation}
    \mathrm{Area}(\mathrm{Type5}) = K^2\times\mathrm{Area}(\mathrm{HondaCRV})
\end{equation}

Let $b_i$, and $b_t$ denote the adjustment to impeller and turbine exit angles, respectively,
\begin{equation}
    \alpha_i(\mathrm{Type5}) = \alpha_i(\mathrm{HondaCRV}) + b_i
\end{equation}

\begin{equation}
    \alpha_t(\mathrm{Type5}) = \alpha_t(\mathrm{HondaCRV}) - b_t
\end{equation}

And $b_i'$, $b_t'$, and $b_s'$  denote the adjustment to impeller, turbine, and stator inlet angles, respectively,

\begin{equation}
    \alpha_i'(\mathrm{Type5}) = \alpha_i'(\mathrm{HondaCRV}) - b_i'
\end{equation}

\begin{equation}
    \alpha_t'(\mathrm{Type5}) = \alpha_t'(\mathrm{HondaCRV}) + b_t'
\end{equation}

\begin{equation}
    \alpha_s'(\mathrm{Type5}) = \alpha_s'(\mathrm{HondaCRV}) + b_s'
\end{equation}
The values of $K$, $b_i$, $b_t$, $b_i'$, $b_t'$, and $b_s'$ that minimize the absolute value of $\Phi$ at the steady state condition of unity torque ratio and unity speed ratio are obtained through a greedy search,

\begin{equation}
    K = 2.73; b_i = 43.0693^{\circ}; b_t = 3.3333^{\circ}; b_i' = 3.5588^{\circ}; b_t' = 0.0980^{\circ}; b_s' = 2.5098^{\circ}
\end{equation}
Accordingly,  Table \ref{tab:my-table} is updated and TC parameters for Type 5 wind turbine are given below,

\begin{table}[h!]
\centering
\caption{Torque Converter Parameters for Type 5 Wind Turbine}
\label{tab:Type5table}
\begin{tabular}{|l|l|l|l|}
\hline
Fluid density ($\rho$)         & 840 kg/$m^3$  & Fluid inertia length ($L_f$)       & 0.7082 m   \\ \hline
Flow area (A)             & 0.0797 $m^2$  & Impeller inertia ($I_i$)           & 0.092 kg $m^2$  \\ \hline
Impeller radius ($R_i$)      & 0.2705 m     & Turbine inertia ($I_t$)            & 0.026 kg$m^2$  \\ \hline
Turbine radius ($R_t$)       & 0.2007 m    & Stator inertia ($I_s$)             & 0.012 kg$m^2$  \\ \hline
Stator radius ($R_s$)        & 0.1815 m    &  Impeller Shock loss coefficient ($C_{sh}$)    & 1.011      \\ \hline
Impeller exit angle ($\alpha_i$)  & 59.3 $ ^{\circ} $ & Turbine Shock loss coefficient ($C_{sh}$)    & 1.8         \\ \hline
Turbine exit angle ($\alpha_t$)   & -56.47 $ ^{\circ} $ &  Stator Shock loss coefficient ($C_{sh}$)    & 0.773      \\ \hline
Stator exit angle ($\alpha_s$)   & Initialize   & Impeller design constant ($S_i$)   & -0.001 $m^2$   \\ \hline
Impeller inlet angle ($\alpha_i'$) & -44.3 $ ^{\circ} $  & Turbine design constant ($S_t$)    & -0.00002 $m^2$ \\ \hline
Turbine inlet angle $\alpha_t'$)  & 59.3 $ ^{\circ} $  & Stator design constant ($S_s$)     & 0.002 $m^2$    \\ \hline
 Stator inlet angle ($\alpha_s'$)         & 62.87 $ ^{\circ} $ & Frictional loss coefficient (f) & 0.197  \\ \hline
\end{tabular}
\end{table}

\begin{figure}[h] 
\centering
\includegraphics[width=0.8\textwidth]{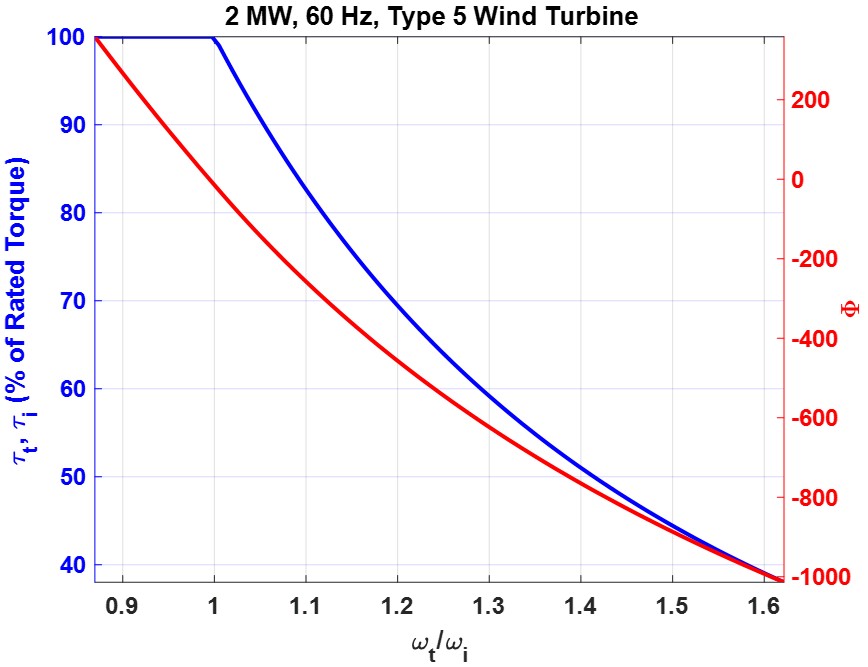}\hspace{2pc}%
 \caption{\label{Fig:PhiZeroCrossing}Variation of $\Phi$ across the speed ratio.}
\end{figure}

For these parameters, the zero crossing of $\Phi$ at unit speed ratio is evident from Figure \ref{Fig:PhiZeroCrossing}. This will help to determine stator exit angle at different steady-state conditions around the unit speed ratio and hence the stator blade control strategy. 

\subsection{Torque Converter Initialization for Type 5 Wind Turbine}
\begin{figure}[h] 
\includegraphics[width=\textwidth]{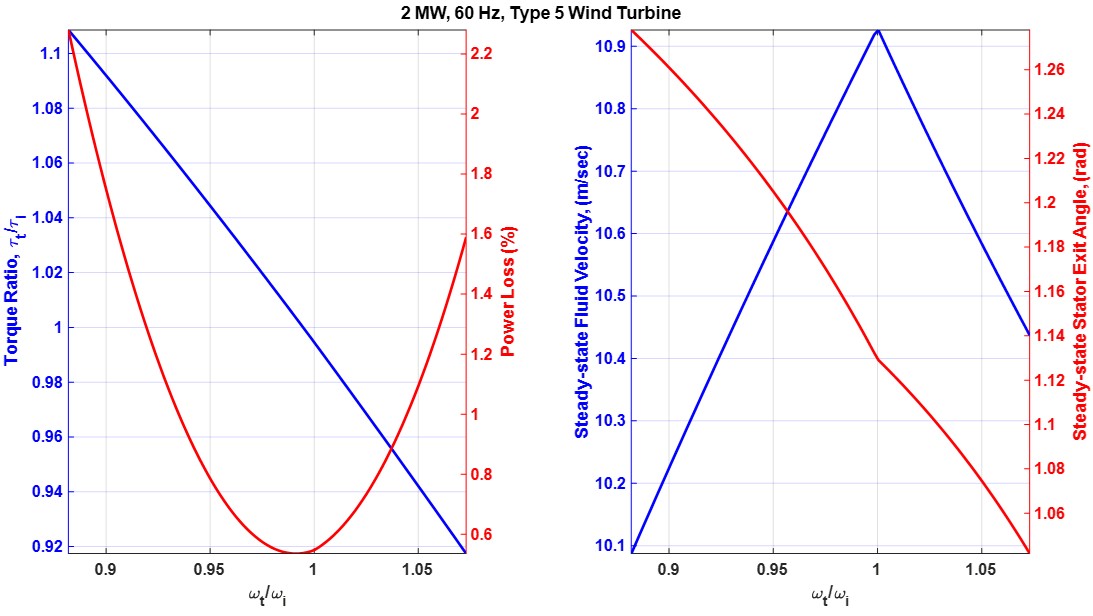}\hspace{2pc}%
 \caption{\label{Fig:Type5Sense}Steady-state sensitivity across the speed ratio.}
\end{figure}

Direction of stator angle change is derived from the TC’s steady-state analysis at $\nu$'s value across the Type 5 wind turbine's operating speed range. Given the $\nu$, and corresponding steady state turbine torque from (\ref{eq:TvsSR}), the steady state value of flow velocity $V_0$, stator exit angle $\alpha_{s0}$, and impeller torque $\tau_i$ are calculated by equating the right hand side to zero for equations (\ref{eq:tto}), (\ref{eq:Phi}), and (\ref{eq:tio}), respectively.  This process is repeated for the entire range of Type 5 Wind turbine specific $\nu$ and Figure \ref{Fig:Type5Sense} is obtained. Notice the reduced range in speed ratio ($0.8819 \leq \nu \leq 1.119$), where a steady-state initialization is feasible for the scaled geometric parameter and adjusted exit angle based Type 5 wind turbine torque converter. The percent power loss is calculated as,
\begin{equation}
    P_{\mathrm{Loss}}(\%) = 100\times \frac{\omega_i\tau_i-\omega_t|\tau_t|}{\omega_t|\tau_t|}
\end{equation}

The decreasing trend in the steady-state stator exit angle will be used in the next section to design the TC turbine side speed governing system.

\subsection{Stator Guide Vane Control}
\begin{figure}[h] 
\centering
\includegraphics[width=0.8\textwidth]{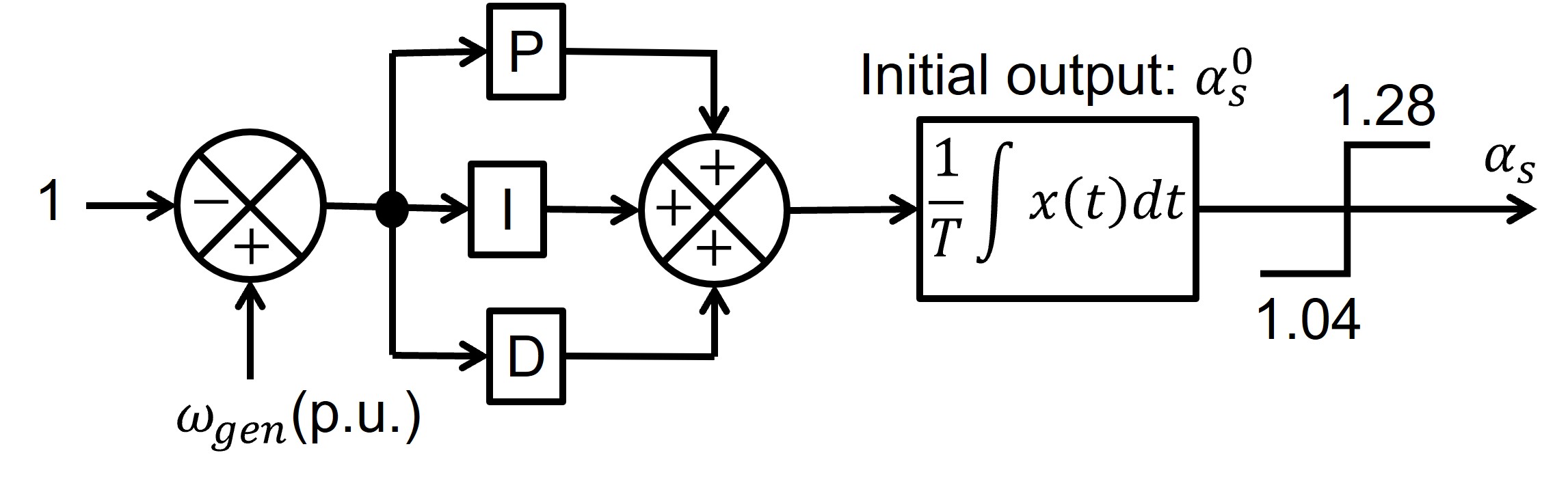}\hspace{2pc}%
 \caption{\label{Fig:Type5PID}PID based Type 5 TC Turbine Speed Governor.}
\end{figure}

The stator blades serve as guide vanes to redirect the hydraulic fluid flow between the impeller and turbine side of the torque converter (TC). Stator angles that measure the guide vane positions are adjusted to correct any speed deviations from the synchronous generator. At any instance, a proportional-integral-derivative (PID) control is applied to either increase or decrease the TC’s stator angle, given the generator speed deviation. From Figure \ref{Fig:Type5Sense}, it is evident that the stator exit angle will need to be decreased (increased) in the case of TC's turbine side under (over) speed. Hence, the PID control will update stator exit angle according to Figure \ref{Fig:Type5PID}. The initial output of the integrator and stator exit angle value range are determined from the analysis reported in Figure \ref{Fig:Type5Sense}. The PID coefficients will be leveraged from \cite{muller_h_grid_2006}, and tuned in future work. 

\section{Conclusions}
Type-5 wind turbines can take advantage of the hydraulic TC constant turbine speed characteristics and hence has a potential to behave like a traditional synchronous generator which can help improve system inertia. This paper describes the validation of an automobile grade TC, its scaling to Type-5 wind turbine specification, and PID control mechanism for the synchronous generator speed governing system. The stator exit angle $\alpha_s$ related to the guide vane position, controls the returning fluid from the turbine to the impeller and it is used as a control variable to maintain generator's synchronous speed. Future work will focus on the PID control tuning, and integration of FAST8 aeroelestic tool. These will enable analyzing
Design Load Cases (DLC) such as normal operation (1.1) and normal operation with low-voltage-ride-through (LVRT). The fully integrated model will be completed on Simulink platform from MathWorks, Inc. \cite{MATLAB:2020a}. After the intellectual property (IP) review, the full dynamic model  will be made available on GitHub.

\ack
Authored by Battelle Energy Alliance, LLC under Contract No. DE-AC07-05ID14517 with the U.S. Department of Energy. Work supported through the U.S. Department of Energy's (DOE) Wind Energy Technologies Office (WETO).

\section*{References}

\bibliographystyle{iopart-num}
\bibliography{./bib/deep_wind_23.bib}

\end{document}